\documentclass[11pt]{article}
\usepackage{graphicx}

\begin{document}
\begin{center}
\Large{\bf The role of dynamics on the habitability of an Earth-like planet}
\bigskip \\
\large{ELKE PILAT-LOHINGER}\bigskip \\
{Institute of Astrophysics, University of Vienna, \\
  T\"urkenschanzstrasse 17, A-1180 Vienna, Austria}
\bigskip \\
\end{center} 

\abstract{
From the numerous detected planets outside the Solar system, no
terrestrial planet comparable to our Earth has been
discovered so far. The search for an Exo-Earth  is certainly a big
challenge which may require the detections of planetary
systems resembling our Solar system in order to find life like on Earth.
However, even if we find  Solar system analogues, it is not
certain that a planet in Earth position will have similar circumstances
as those of Earth. Small  changes in the architecture of the giant planets
can lead to orbital perturbations which may change the conditions of habitability for
a terrestrial planet in the habitable zone (HZ). We present a numerical 
investigation where we first study the motion of test-planets in a particular 
Jupiter-Saturn configuration for which we can expect strong gravitational 
perturbations on  the motion at Earth position according to a previous
work. In this study, we show that these strong perturbations can be
reduced significantly by the neighboring planets of Earth.
In the second part of our study we investigate the motion of
test-planets in inclined Jupiter-Saturn systems where we analyze 
changes in the dynamical behavior of the inner planetary system.
Moderate values of inclination seem to counteract the
perturbations in the HZ while high inclinations induce more chaos
in this region. Finally, we carry out a stability study of the actual orbits of
Venus, Earth and Mars moving in the inclined Jupiter-Saturn systems for
which we used the Solar system parameters. This study shows that the three
terrestrial planets will only move in low-eccentric orbits if Saturn's
inclination is $\leq 10^o$. Therefore, it seems that it is advantageous for
the habitability of Earth when all planets move nearly in the same plane.  



\section{Introduction}

The search for planets outside the Solar system has shown an
unexpected diversity of planetary systems. Strange new worlds like 
pulsar planets, hot-Jupiters and high eccentricity motion of planets as well as
planets in binary star systems were not expected by astronomers  when starting
their search  for other worlds. The diversity of planetary systems is evidence
that phases of instability during the formation process have shaped these
systems.  Numerical simulations, which showed hypothetical scenarios for
the evolution of the Solar system -- like  e.g.\ the Nice models 
(Tsiganis et al.\ 2003, Morbidelli et al.\ 2009, Levison et al.\ 2011) --
indicated chaos in the orbital motion of the outer planetary system before the
final architecture was reached.
The strength and duration of this turbulent phase  could be
important for the future evolution of a planetary system, as
different  evolutionary tracks for the planetary motion will lead to
different final architectures of a system. 

The fact that the Solar system planets move in nearly the same
  plane and  in nearly circular orbits could allude that the instability
  phase in our planetary system was moderate and, therefore, advantageous for
  the habitability of our Earth which is still the only planet known where
  life as we know it, exists. Taking into account that complex life on Earth needed
  an evolution over a very long time, we assume that the low-eccentricity
  and quite stable planetary motion might be a necessary requirement for
  ``Earth-like'' habitability. Therefore, the planet has to move in the
so-called habitable zone (HZ) which is the region around the star where liquid
water is stable on the surface of an Earth-type planet 
(Kasting et al.\ 1993)\footnote{Besides this basic requirement many other
  conditions of astrophysical, geophysical, chemical and biological nature
  have to be fulfilled that a planet can be considered as habitable planet.}. 
In this study, we assume that the planet has one Earth ocean of surface water, 
where the carbonate-silicate cycle controls the CO$_2$ level in equilibrium
with a temperature above the freezing in the HZ. 

Therefore, we define the boundaries of the HZ to be at 0.95 au and 1.37 au,  
where the inner border is determined by the runaway greenhouse effect according 
to the work by Leconte et al.\ (2013) and the outer boundary is taken from Kasting et
al.\ (1993)\footnote{This outer boundary does not take
into account CO$_2$ clouds which can significantly affect the
temperature-CO$_2$ coupling. These effects may shift 
the outer boundary to 1.7 au or 2 au (see e.g.\ Forget \& Pierrehumbert, 1997
or Mischna et al., 2000). }.  We studied the dynamical behavior in this area
via long-term computations of a set of test-planets moving in
Sun-Jupiter-Saturn like configurations.
Numerical stability studies of planetary motion in the HZ have been
carried out since the detection of the first extra-solar planet by
Mayor \& Queloz in 1995.  These studies examined the stability or formation of  
certain detected extra-solar planetary systems in single and multiple star systems. The first
relevant publications are e.g.\ Gehman et al.\ (1996), Jones \& Sleep (2002),
Menou \& Tabachnik (2003), Jones et al.\ (2005) and many
others\footnote{Unfortunately the  
  literature is too rich on this subject to cite all relevant articles here.}
(see e.g.\  S\'andor et al.\ (2007), Eggl et al.\ (2012, 2013), M\"uller \&
Haghighipour (2014) ) whose general numerical studies can be applied
to many of the discovered systems. 

The present numerical study was motivated by previous investigations of 
Pilat-Lohinger et al.\ (2008a\& b -- hereafter PL08a and PL08b) which showed that
 strong secular perturbations act on the motion of test-planets in the
 HZ  which can lead to high eccentricities for certain orbits 
(see the arched band in Fig.\ 2 in PL08a).  
Here, we focus on the region where secular
perturbations affect the motion of an Earth-like planet at 1 au. This is the
case  when Saturn's semi-major axis is 8.7 au. Then the  
motion of a terrestrial planet at 1 au would vary from nearly
circular to highly eccentric (with an eccentricity $\geq 0.6$). Consequently, 
the orbit will either be entirely, mostly, or only in average
within the HZ.\footnote{This is similar to the classification of different HZs 
for binary star systems  where between (i) permanent,
(ii) extended  and (iii) averaged HZs is distinguished (see Eggl et al.,
2012). } An orbit with eccentricity $\geq 0.6$ would certainly belong to the latter
case which would lead to strong variations in the surface temperature
as shown by Williams \& Pollard (2002) for  Earth. In their study, these authors
increased the eccentricity of Earth up to 0.7  and showed that 
Earth can stay habitable as the planet did not
freeze out at its aphelion distance, and no complete water evaporation
occurred  at its perihelion distance. 

Because the studies PL08a and PL08b have shown that the dynamical
  behavior in the HZ can change significantly when we modify the architecture
  of the system, we checked the influence of the ice-planets Uranus
  and Neptune  and of the neighboring planets Venus and Mars in this particular
  Jupiter-Saturn system. Our study shows that especially Venus plays an
  important role (as it was also found in PL08a). 

In the second part of this study, we considered inclined
  Jupiter-Saturn configurations for which we increased Saturn's inclination up
  to 50$^\circ$. The numerical simulations showed that systems with inclinations
  between 10$^\circ$ and 40$^\circ$ indicate a significant decrease in the maximum
  eccentricity for the test-planets in the HZ. And for higher inclinations the
chaotic area increased. 

Finally, we were interested in the dynamical behavior  of the
  terrestrial planets Venus, Earth and  Mars when moving in the inclined
  Jupiter-Saturn systems. These computations showed chaotic perturbations and an
  escape of Mars for an inclination of $20^\circ$ for Saturn's orbit. 
For higher inclinations, the orbit of Mars immediately becomes
  chaotic  and perturbs the motion of Earth and
  Venus, as well. Moreover, the chaotic area increases with Saturn's inclination. 

This paper is organized as follows. We describe the perturbations in a 
planetary system in section 2, and present the computations in section 3. 
The study in a particular planar Jupiter-Saturn system is shown in
section 4, and in the inclined Jupiter-Saturn configurations in section 5. 
Finally, in section 6, we examine the motion of the actual orbits of Venus,
Earth and Mars in the inclined Jupiter-Saturn configurations.

\section{Perturbations in a Planetary System}

Planetary orbits are described by a set of orbital elements where the
semi-major axis $a$ and the eccentricity $e$ define the size and the shape of the 
orbit and the angles: inclination $i$, argument of perihelion $\omega$ and
longitude of the ascending node $\Omega$  specify the
orientation of the orbit in space. Finally, the mean anomaly $M$ defines the
orbital position of the body. As soon as more than one planet
is orbiting a star, there will be a variation of these orbital parameters due
to gravitational interactions between the planets. 
From studies of the Solar system, we know that resonant perturbations may
influence the orbital motion significantly. For our study the  mean motion
resonances (also known as orbital resonances) and 
secular resonances are of special interest. 

\underline{Mean motion resonances (MMRs)} occur when the orbital
periods of two celestial bodies are close to a ratio of small integers

\begin{equation}
\frac{n_1}{n_2} \sim \frac{k_1}{k_2}\,,
\end{equation}   

\noindent
where $n_1$ and $n_2$ are the mean motions (={\it Orbital Period/$2 \pi$}) of
the celestial bodies and $k_1, k_2$ are integers. Orbital resonances are the source of 
both stability and chaos, depending sensitively upon parameters and initial conditions. 
Well known examples of MMRs in the Solar system are the so-called Kirkwook gaps in the 
asteroid belt. The locations of these gaps correspond to  MMRs with Jupiter.

\underline{Secular resonances (SR)} occur when one of the precession
frequencies of a celestial body -- related to the motion of $\omega$  or
$\Omega$ 
-- is equal (or a linear combination) of the proper modes of the planetary
motion ($g_l, s_l$ with
$l=1, \ldots N$ where $N$ is the number of planets -- for the
Solar System $l=1$ corresponds to Mercury $\ldots$
$l=8$ is Neptune). When considering the giant planets
Jupiter and Saturn moving on low inclination and low eccentricity orbits,
these  frequencies can be  deduced by the following  
secular linear approximation (see e.g. Murray \& Dermott 1999)

\begin{eqnarray}
g &=&  {n\over{4}}  \left (  {{m_{J}}\over{m_{\rm Sun}}} \alpha_{J}^2 b^{(1)}_{3/2}(\alpha_{J})  
+ {{m_{S}}\over{m_{\rm Sun}}} \alpha_{S}^2 b^{(1)}_{3/2}(\alpha_{S})\,,
\right ) \\
s &=& -g  \nonumber
\label{epl-eq2}
\end{eqnarray}

\noindent
where $\alpha_{J} = a/a_{J}$, $\alpha_{S} = a/a_{S}$ with 
$a_J, a_S, a_{TP}$  being the semi-major axes of Jupiter, Saturn and the test-planet, 
respectively. The quantities $m_J$ and $m_S$ are
the masses of Jupiter and Saturn, $m_{\rm Sun}$ is the mass of the Sun 
  and  $b^{(1)}_{3/2}$ is a Laplace coefficient. Moreover, the test-planets
  are considered to be mass-less compared to the other masses and they
must have nearly zero eccentricities and inclinations. 
Solutions of Eq.~2 for $g(a_{TP}, a_S)  = g_5(a_S)$ and
$g(a_{TP}, a_S) = g_6(a_S)$ are SRs connected to Jupiter or Saturn, respectively. 

It is well known that SR and low order MMRs may lead to significant changes
in the orbital motion of bodies with large variations in their eccentricities. This could
cause problems for a  planet in the HZ, which is a quite small region. If
this planet moves in a high eccentricity orbit, it will leave
the HZ periodically, which might change the conditions for it habitability.

\section{Dynamical Model and Computations}

To study the dynamics of test-planets in the HZ  between 0.95 au and 1.37 au,
we used the restricted problem which is commonly used for such
investigations.  The test-planet is
considered to be mass-less and  move in the gravitational field
of the Sun, Jupiter and Saturn without perturbing their orbits. 

In the first part of this study, Jupiter was fixed to its actual orbit at 5.2 au and 
Saturn's semi-major axis was changed to 8.7 au  while the other orbital elements
were those of the Solar system (as published in PL08a, Table 1). 
For the integrations,
the Bulirsch-Stoer integration method was used and the stability of the
orbital motion was verified calculating  the Fast Lyapunov
Indicator (FLI). This chaos indicator was introduced by Froschl\'e et
al.\ (1997) and is based on the Lyapunov Characteristic Exponent (LCE) (see
e.g.\ Froeschl\'e, 1984). When calculating the FLI, one can
easily distinguish between regular and chaotic motion due to the
growth of the largest tangent vector of the dynamical flow. The growth can be
either linear or exponential where the latter characterizes chaotic
motion. 

In the second part, we studied inclined Jupiter-Saturn systems
for which we used  the hybrid
integration method of the Mercury6 package (Chambers, 1999). 
For the computation of the maximum eccentricity maps we took again the initial
conditions published in  PL08a  and varied 
the inclination of Saturn from $10^\circ$ to $50^\circ$  in steps of $10^\circ$. 

Finally, in the third part we examined the stability of the
terrestrial planets (Venus to Mars) in the different inclined Jupiter-Saturn
systems for which we used the orbital parameters of the Solar system.

\section{The planar Jupiter-Saturn system}

\begin{figure}
\centering 
\includegraphics[height=7cm,angle=270]{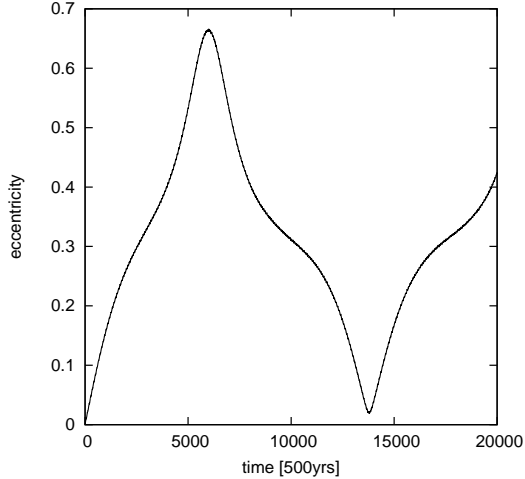}    
\caption[]{Time evolution of the eccentricity of a test-planet at 1 au for
  Saturn at 8.7 au.}
\label{epl-f1}
\end{figure}
 
In PL08a, the locations of the two main secular
frequencies associated with the precession of perihelia of Jupiter and
Saturn (known as $g_5$ and $g_6$ frequencies in the Solar system) were
calculated using the  
frequency analysis of Laskar (1990; see also Robutel \& Gabern 2006). 
The result of $g = g_5$ (when applying Eq.~2) was found to be in good
agreement with the arched band of higher eccentricity shown in Fig.\ 2 of PL08a.
Due to this perturbation, the motion of a test-planet at 1 au
indicates strong variations in eccentricity if
Saturn orbits the Sun at 8.7 au. 
In Fig.~1, one can see a long periodic variation of the eccentricity
which changes from circular to highly eccentric ($e > 0.6$) within 3 Myrs. Taking
0.95 au and 1.37 au as HZ boundaries, the entire orbit of this planet is in the HZ for only 
the first and last 150000 yrs of the 7 Myrs. For $0.05 < e < 0.4$, the
planet is in the HZ for most of the time and leaves this zone only  at
peri-center. In case of $e > 0.4$, the planet exits the HZ at peri- and
apo-center as it can be seen in Fig.~\ref{epl-f2}. In this figure, one can
see that less than 23\% of the orbit
of a planet with eccentricity of 0.7 (black line) would be in the HZ.
However, a study by William and Pollard (2002) showed that Earth 
remains habitable  
even for such a high eccentricity. Of course, there would be strong
variations of the surface temperature and then, probably the evolution of life
would have been different to ours.
If we shift the outer border to 1.67 au\footnote{When taking the maximum
greenhouse  effect as limit} then the planet at 1 au leaves the HZ only when approaching 
the peri-center for all eccentricities $> 0.05$ and about 70\% of a highly
eccentric orbit ($e=0.7$) would be in the HZ.

\begin{figure}[h]
\centering 
\includegraphics[height=9cm,angle=270]{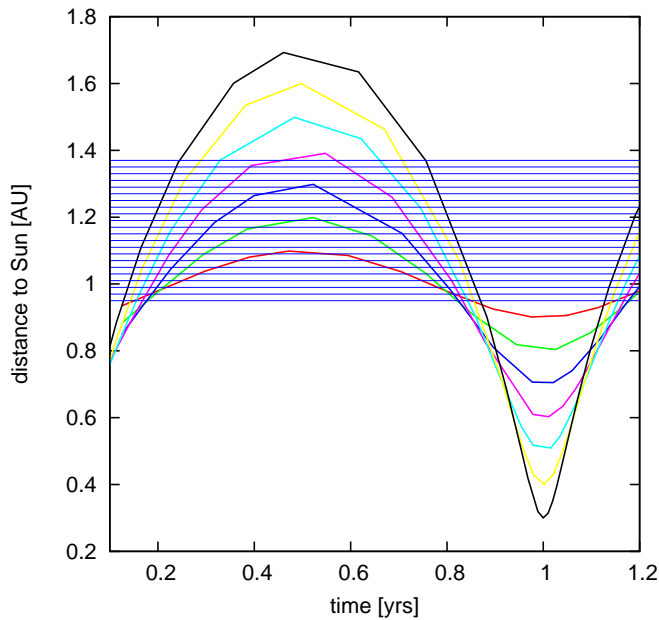}    
\caption[]{Distance to the Sun of a planet at 1 au. Each curve shows the
  variation of the distance over one orbital period of the planet using
  different orbital eccentricities represented by the different colors: 0.1
  (red), 0.2 (green), 0.3: (blue), 0.4 (magenta), 0.5 (light blue), 0.6
  (yellow) and 0.7 (black). The blue area labels the HZ.}
\label{epl-f2}
\end{figure}

Regardless of the choice of the HZ borders we recognized a difference
between orbits 
with moderate ($e < 0.4$) and high eccentricities ($e \geq 0.4$) from the
dynamical point of view. 
For the latter, small perturbations occur in semi-major axis ($a$),
inclination ($i$) and ascending node ($\Omega$) (see
Figs.~\ref{epl-f3} a-c) whenever the planet's eccentricity exceeds 0.4. The
strongest fluctuations appear when the orbit reaches its maximum eccentricity
after 3 Myrs (see also Fig.~\ref{epl-f1}). Moreover, the evolution of the
planet's  node (Fig.~\ref{epl-f3} bottom) shows transitions
between rotation and libration when $e > 0.4$, which is known to be an indication of chaos.

\begin{figure}[]
\includegraphics[height=17cm,angle=270]{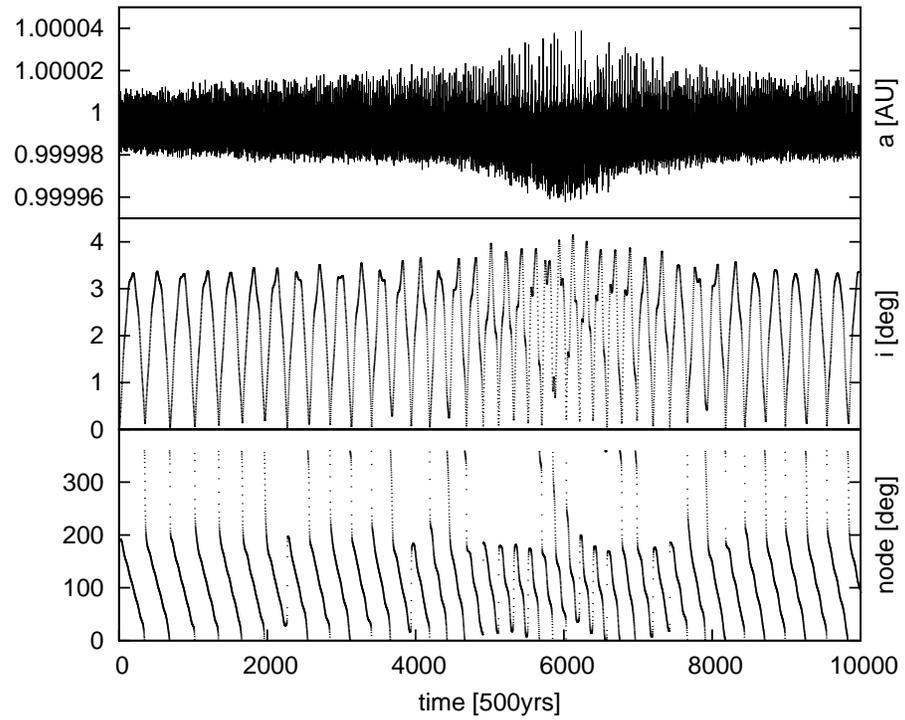}  
\caption[]{Time evolution of the semi-major axis $a$ (top), the
  inclination $i$ (middle) and
  the node (bottom) of the planetary orbit when its eccentricity
  exceeds 0.4. One can see small fluctuations in the top and middle panels and
  transitions between libration and circulation in the bottom panel.} 
\label{epl-f3}
\end{figure}

To verify this orbital behavior, a long-term computation of the system over 100
Myrs has been performed where we  also
calculated the Fast Lyapunov Indicator (FLI) to determine the
dynamical state of the orbit via this chaos indicator.
In Fig.~\ref{epl-f4}, one can see that the continuation of the simulation
  of Fig.~\ref{epl-f1}  to longer times immediately points to a
long period of high-eccentricity motion. More precisely, for more than 42 Myrs, the planet's
eccentricity is almost always in the range $\geq 0.4$, where chaos may arise.
This can be seen clearly in  Fig.~\ref{epl-f5} where the FLI of
the planetary orbit increases  whenever the eccentricity reaches
its maximum value. This leads to a step-like increase  of the FLI. 
Between 12 Myrs and 55 Myrs, when the eccentricity is almost always $\geq 0.4$,
the curve has a steeper rise.

\begin{figure}
\centering 
\includegraphics[height=9cm,angle=270]{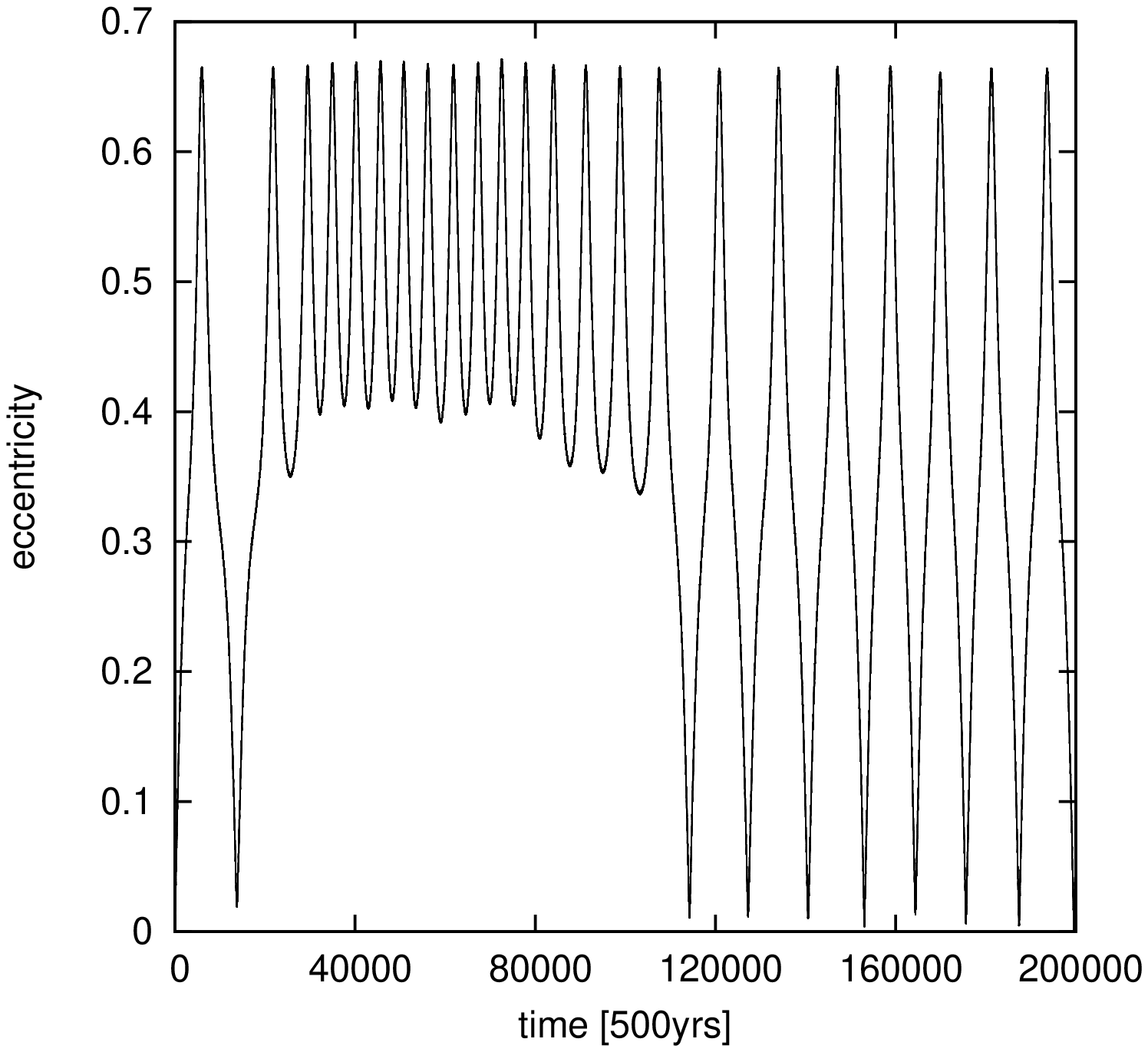}    
\caption[]{Same as Fig.~\ref{epl-f1} but for 100 Myrs} 
\label{epl-f4}
\end{figure}
\begin{figure}
\centering 
\includegraphics[height=9cm,angle=270]{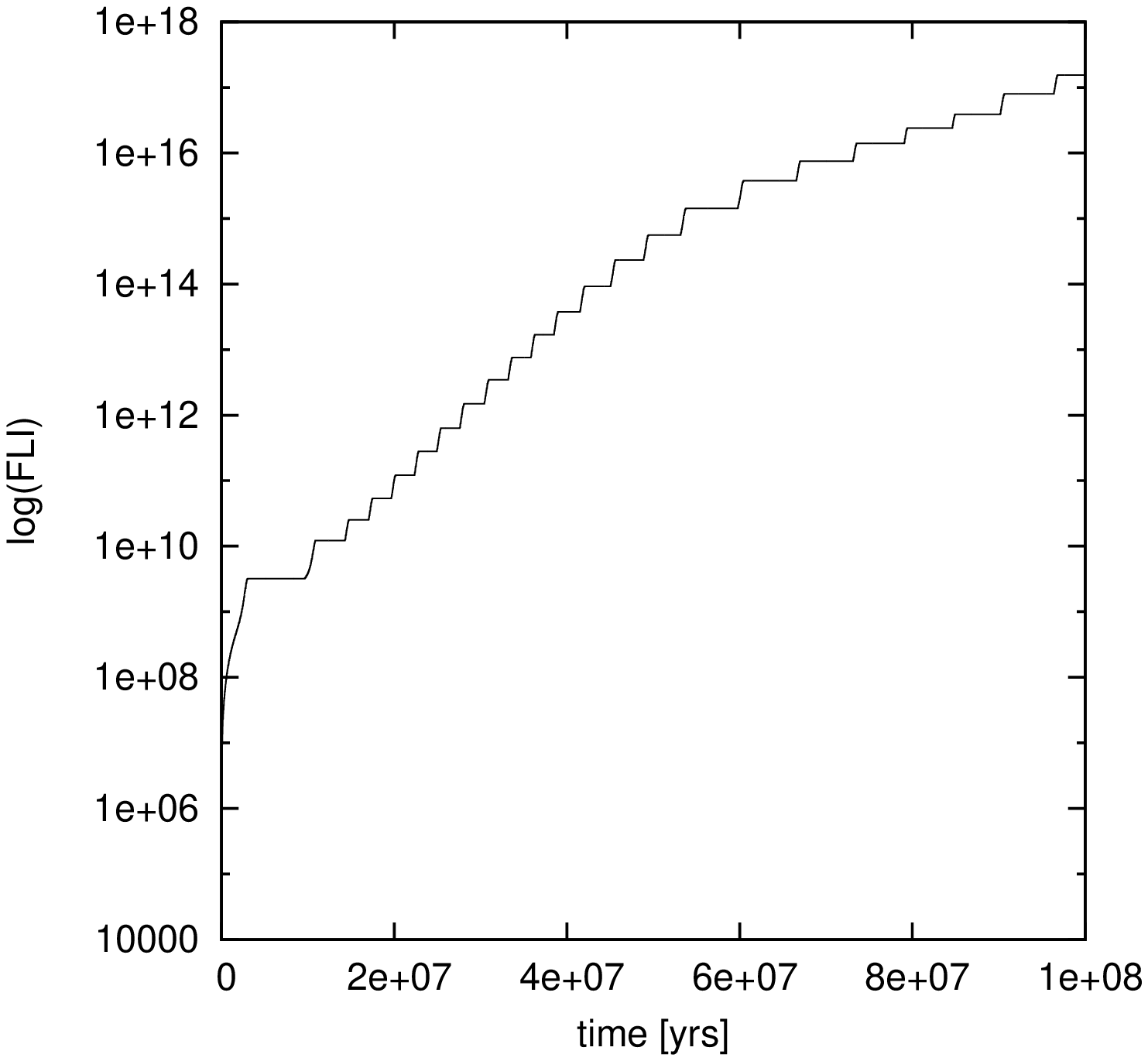}    
\caption[]{FLI stability plot for a test-planet at 1 au.} 
\label{epl-f5}
\end{figure}

It is important to note that this study was carried out for a particular Jupiter-Saturn
configuration. 
To generalize the results,  we computed the orbital evolution using different
relative positions of Jupiter and Saturn by varying Saturn's mean
anomaly and the argument of peri-center  (see Fig.~\ref{epl-f6}).
As an indicator for dynamical habitability, we used the maximum
eccentricity (max-e) which shows whether peri- and apo-center distance
are in the HZ. Together with the evolution of the eccentricity, we can estimate 
if the orbit is permanently, mostly, or in average in the HZ  (which depends
also on the choice of the HZ borders). Moreover, the eccentricity
indicates changes due to secular perturbations as shown in PL08a and PL08b.
Fig.~\ref{epl-f6} shows the max-e values of a test-planet at 1 AU 
for different starting positions of Saturn (x-axis) and different orientations
of Saturn's orbit (y-axis). The color purple shows constellations of lowest
max-e values and yellow corresponds to highest values of
max-e. According to this map,  one can see that the
different Jupiter-Saturn configurations indicate mainly high 
values of maximum eccentricities (i.e.\ red, orange and yellow areas) and
only a few configurations show moderate 
values of max-e between 0.2 and 0.35. This is the case when the relative
values of both, the mean motions and the peri-centers are around $180^\circ$. 
The latter can also be zero according to Fig.~\ref{epl-f6}.
Only for these few combinations of the giant planets (i.e.\ blue and
  purple squares in  Fig.~\ref{epl-f6}), the test-planet at 1 au
has low eccentricities so that the conditions for 
habitability could be quite similar to that of our Earth. 

\begin{figure}[h]
\centering 
\includegraphics[height=9cm,angle=270]{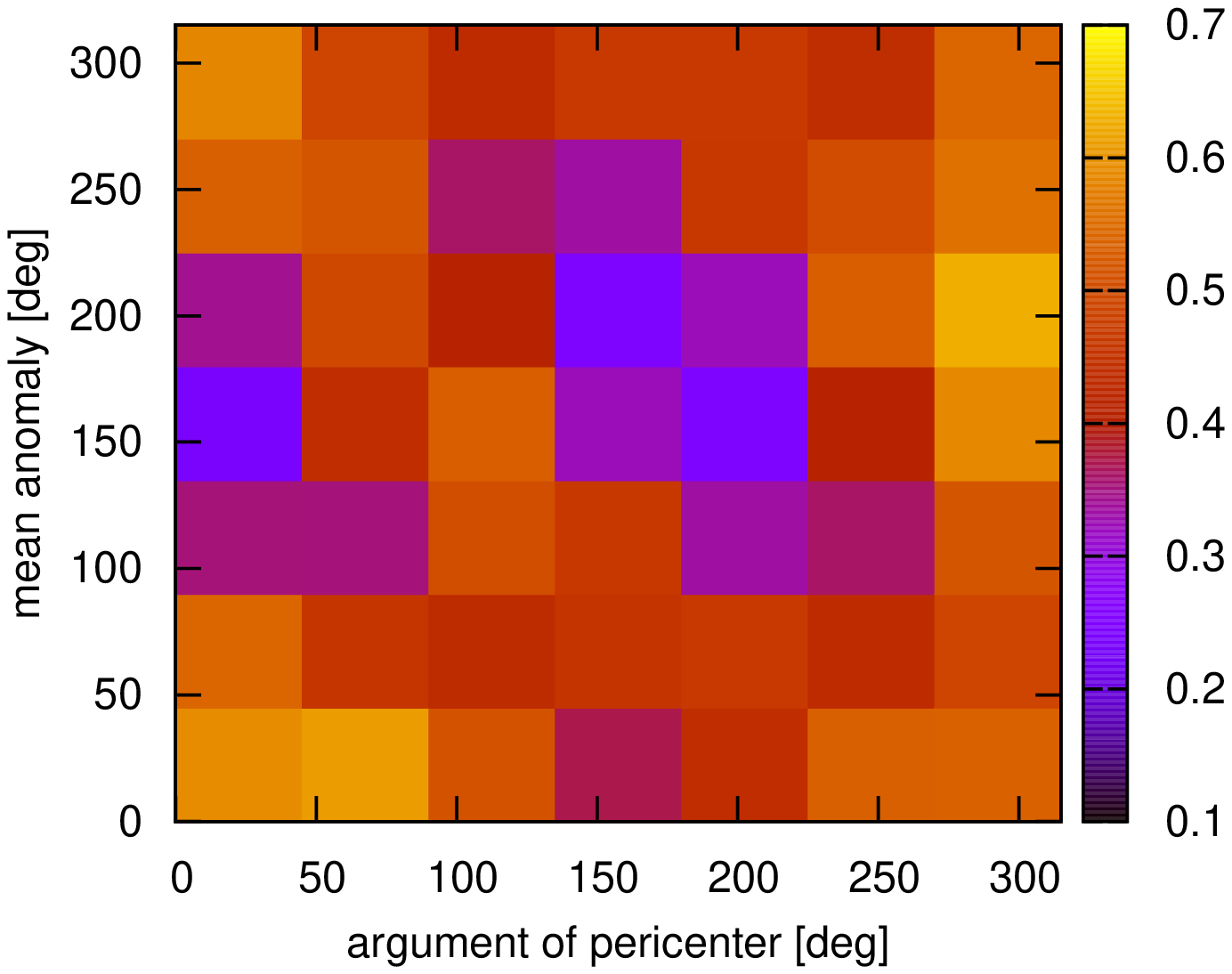}    
\caption[]{Maximum eccentricity of a test-planet at 1 au for various Jupiter-Saturn
  configurations. $a_{Saturn}$ was fixed to 8.7 AU and its relative position
  to Jupiter was changed by varying Saturn's mean anomaly and argument of
  peri-center. Different colors indicate different values of max-e for the
  test-planet at 1 AU (according the color scale).}
\label{epl-f6}
\end{figure}

The planetary system studied here is much simpler
than the Solar system as we took into account only two giant planets. 
For a better comparison with the Solar system, we studied the
influence of Uranus, Neptune and the
neighboring planets of Earth by including these planets in the dynamical model
step by step. The results are shown in Fig.~\ref{epl-f7} where the red
line represents the evolution of the test-planet's eccentricity when
using Jupiter through Neptune as dynamical model. The two ice
planets do not decrease the maximum eccentricity significantly as the signal
shows still a periodic variation in eccentricity (between 0 and 0.6) but with a
different  period. If we add Mars to the system, the
test-planet's eccentricity will be reduced to $< 0.3$ (see the green
line in Fig.~\ref{epl-f7}). However, the most important effect can be
recognized when we include Venus. This planet damps the
eccentricity of the test-planet at 1 au down to nearly circular motion with only
small fluctuation and no secular variations (blue line in Fig.~\ref{epl-f7}). 
This result shows again the important interplay of Venus and Earth which was also 
found in PL08a, where the presence of Earth helped to decrease Venus'eccentricity 
to the observed value. 

It is known that the orbits of Venus and Earth are connected due to a high order MMR (for 
details see e.g.\ Bazso et al., 2010). Nevertheless, it is questionable if  
the 13:8 MMR is also important for the damping of the  eccentricity in the area
affected by the secular perturbation. To check this, we performed similar
computations as shown in Fig.~\ref{epl-f7} for test-planets between 0.9
and 1.1 au, and compared the max-e values of the different systems. The results are
summarized in Fig.~\ref{epl-f8} where we see the secular perturbation in the
Jupiter-Saturn system mainly between 0.96 and 1 au (dotted line with open
squares). If we add Uranus and Neptune to the system, we can see a slight decrease
in max-e for all semi-major axes $\leq 1$ au (see the red line in
Fig.~\ref{epl-f8}). The green line, which represents the result when Mars is
also included, shows clearly that Mars shifts the secular perturbation away from 1
au and consequently,  the maximum eccentricity decreases at this position.
Moreover, we note a significant
decrease in eccentricity in this dynamical model for a planet at 0.96
au. This is the only position where the influence of Mars is stronger than
that of Venus. At this distance, the test planet is
in 2:1 MMR with Mars. This low order MMR counteracts the
secular perturbation so that the eccentricity remains low. For all other semi-major
axes, we observe a strong decrease in eccentricity due to Venus.

\begin{figure}
\centering 
\includegraphics[height=9cm,angle=270]{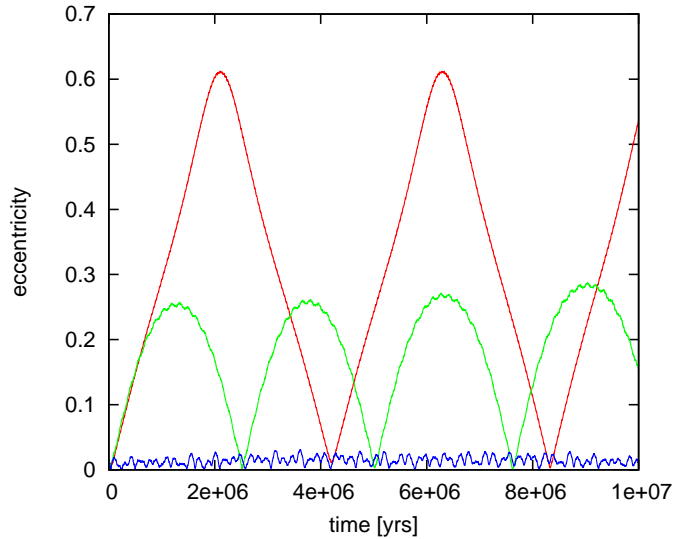}    
\caption[]{Evolution of the eccentricity for an orbit at 1 au in different
  dynamical systems.
}
\label{epl-f7}
\end{figure}

\begin{figure}
\centering 
\includegraphics[height=9cm,angle=270]{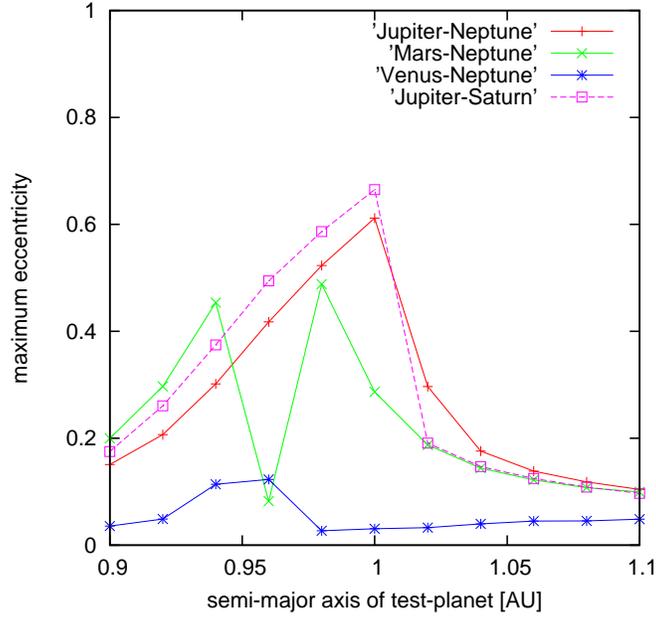}    
\caption[]{Maximum eccentricity for orbits in the HZ using different
  dynamical systems.
}
\label{epl-f8}
\end{figure}

The slight increase of max-e  in the Venus-Neptune system between 0.94 and
0.96 can be explained by the interaction of two important MMRs  which
influence this area. These are the 2:1 MMR with Mars at 0.96 au and the 3:2
MMR with Venus at  0.948 au. In case where these MMRs overlap, chaos will arise 
which can be confirmed by FLI computations. Nevertheless, this might not exclude 
regular motion over long time-scales similar to the case of the
Solar system which is chaotic but the planetary motion is 
stable for the life-time (i.e.\ the time on the main-sequence) of the
Sun.

Even if we take our Solar system as reference system for this study, where the
planets move in nearly the same plane, we should not ignore the possibility of
mutually inclined planetary orbits.

\begin{figure}
\centering{
\includegraphics[height=8.5cm]{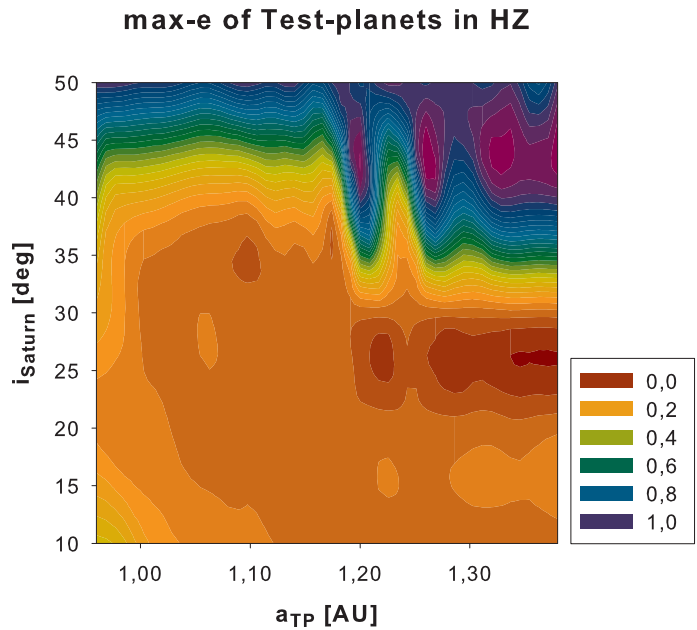}
}
\caption[]{Max-e plot for test-planets in the HZ from 0.95 au
  to 1.37 au for different inclinations of Saturn (y-axis). The maximum
  eccentricity is given by the color code. } 
\label{epl-f9}
\end{figure}

\section{Inclined Jupiter-Saturn systems}
The study of a particular Jupiter-Saturn configuration (when
$a_{\rm Saturn}=8.7$ au) for 
different inclinations of Saturn's orbit (from $10^\circ$ to $50^\circ$) shows the
perturbations of the two giant planets on the motion in the HZ, and that it 
depends on the mutual inclination of Jupiter and Saturn.
The results of the computations are
summarized in Fig.~\ref{epl-f9}, where the maximum eccentricity is plotted for
test-planets between 0.95 au  and 1.37 au. Surprisingly, one can see
low max-e values in the entire HZ  for Saturn’s inclination up to $30^\circ$.
For inclinations larger han that,  
the outer part of the HZ shows quite high eccentricities (green and blue
area) as well as unstable areas (purple regions), while the inner part of the
HZ ($a_{TP} < 1.2$ au) still shows  low max-e values up to an inclination of
$40^o$.  If $i_{\rm Saturn}> 45^\circ$ all orbits in the HZ have maximum
eccentricities $ \geq 0.8$. From this result, we follow that ``Earth-like
habitability'' can be excluded for  highly inclined  Jupiter-Saturn
systems (when $a_{\rm Saturn}=8.7$ au).  

For a  test-planet at 1 au, we observe the following dynamical behavior, \\ 
(i) The nearly planar Jupiter-Saturn system shows  strong variations of the  
eccentricity with a maximum value  $> 0.6$ (see Fig.~\ref{epl-f1}), \\
(ii) For Saturn's inclination from  $10^\circ$ to $40^\circ$ we notice a significant
decrease in max-e for this planet to values around 0.2, and \\
(iii) And high inclinations of Saturn  ($ > 45^o$)  lead again to high 
eccentricities of 0.8 for the test planet.

To get an idea about the variation of the dynamical structure, especially 
in the region of the
significant bend, where higher eccentricity motion occurs due to secular
perturbations of Jupiter and Saturn (as found in PL08a), we studied 
a larger region of the parameter space 
where the semi-major axis of the test-planet is varied between 0.6 to
1.6 au and that of Saturn changes between 8.2 and 9.8 au. The maps
of Figs.~\ref{epl-f10}(a-f) show the 
perturbations of the different Jupiter-Saturn configurations. 
Fig.~\ref{epl-f10}a (top left panel) displays the result corresponding to the 
Solar system parameters. The results of the computations for inclined systems
are shown in Figs.~\ref{epl-f10}b (for $10^\circ$) to \ref{epl-f10}f (for $50^\circ$). 
Pointing our attention to the bend in Fig.~\ref{epl-f10}a, which is
visible up to an inclination of
$30^\circ$, we observe no significant change in the shape up to $i_{Saturn}
=20^\circ$ whereas for $i_{\rm Saturn}=30^\circ$ (right panel in the middle), we recognize stronger
perturbations due to the inclined orbit of Saturn  and a change of the
dynamical structure in this map. A further increase of
Saturn's inclination (bottom left panel) shows last remnants of the bend
between the two horizontal blue stripes at about 8.3 au and 8.9 au, which
are the locations of the 2:1 MMR and the 9:4 MMRs between Jupiter and Saturn.
A comparison of the different figures shows that the spots of high
eccentricity motion 
in the bend (see the blue-green islands in the top left panel) 
change significantly  when increasing
Saturn's inclination.  Like 
 in Fig.~\ref{epl-f9} for $a_{\rm Saturn}=8.7$ au, the maximum
eccentricity will decrease when increasing Saturn's inclination . Since these
perturbations result from a secular frequency associated with the precession of the
perihelion of Jupiter ($g_5$ frequency in the Solar system), it seems that a
mutual inclination of the two giant planets reduces this perturbation. 

In contrast, it can be seen that the perturbations
at positions of MMRs between Jupiter and Saturn are stronger in the
inclined systems as the
horizontal stripes in the figures indicate higher eccentricities.
In the map for $i_{\rm Saturn}=10^\circ$ (top right panel), one can see these
perturbations at about 9.6 au and  8.3 au  corresponding to the
5:2 MMR and 2:1 MMR, respectively.  
Both stripes show a spot-like structure  where the strongest
perturbations are visible for test-planets with semi-major axes $> 1.4$ au when
Jupiter and Saturn are in 2:1 MMR. An increase of  $i_{\rm Saturn}$ to
$20^\circ$ leads to even stronger 
perturbations in this area where all test-planets with $a_{TP} > 1.3$ au
escape (purple stripe) while in the map for $i_{\rm Saturn}= 30^\circ$ (right
middle panel) these test-planets 
indicate lower max-e values so that fewer orbits escape than for $i_{\rm Saturn}= 20^\circ$.  
The max-e map for $i_{\rm Saturn}=40^o$ does not show the perturbation at 9.6 au
anymore. One can see that the dynamical structure has completely changed as
the outer HZ becomes unstable (purple area) for all Jupiter-Saturn
configurations. Moreover, we recognize two stripes of high eccentricities even
in the inner HZ when $a_{\rm Saturn} = 8.3$ au or 8.9 au.
As can be seen from different panels,
the unstable region in the outer HZ increases with Saturn's inclination
indicating that the perturbation is generated by Saturn.
   
\begin{figure}
\centering{ 
\includegraphics[height=5cm]{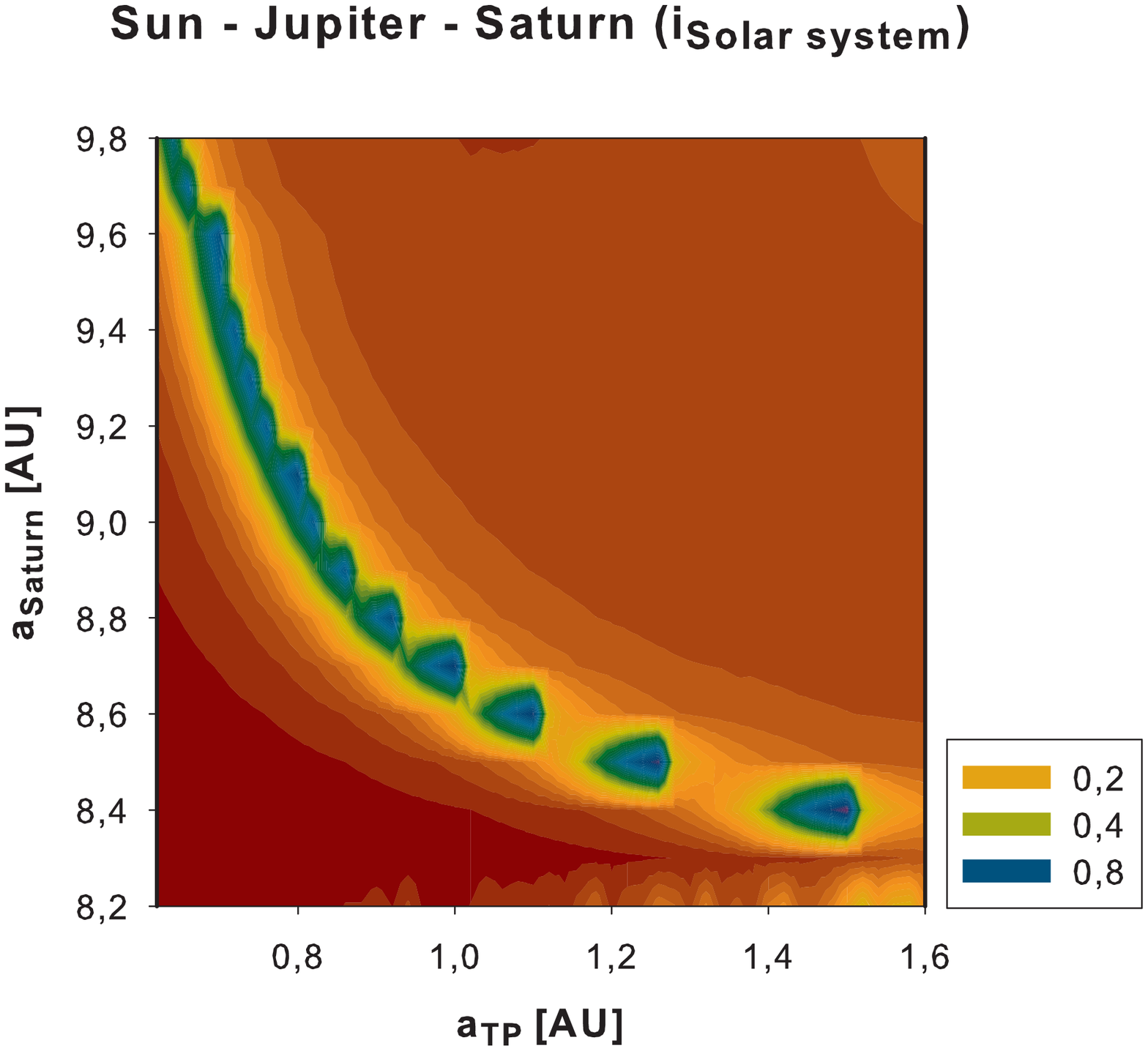}   
\includegraphics[height=5cm]{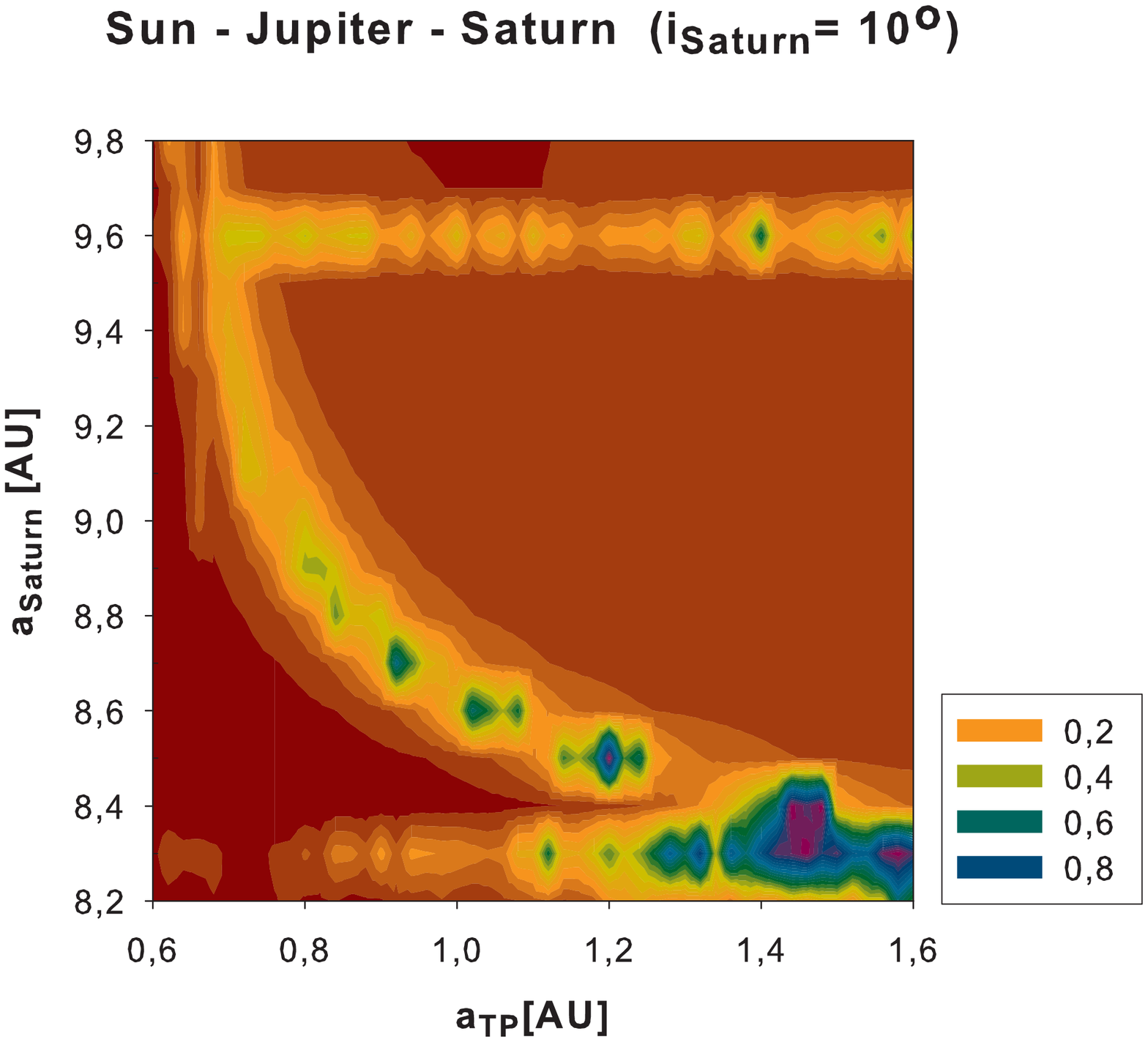} 
\includegraphics[height=5cm]{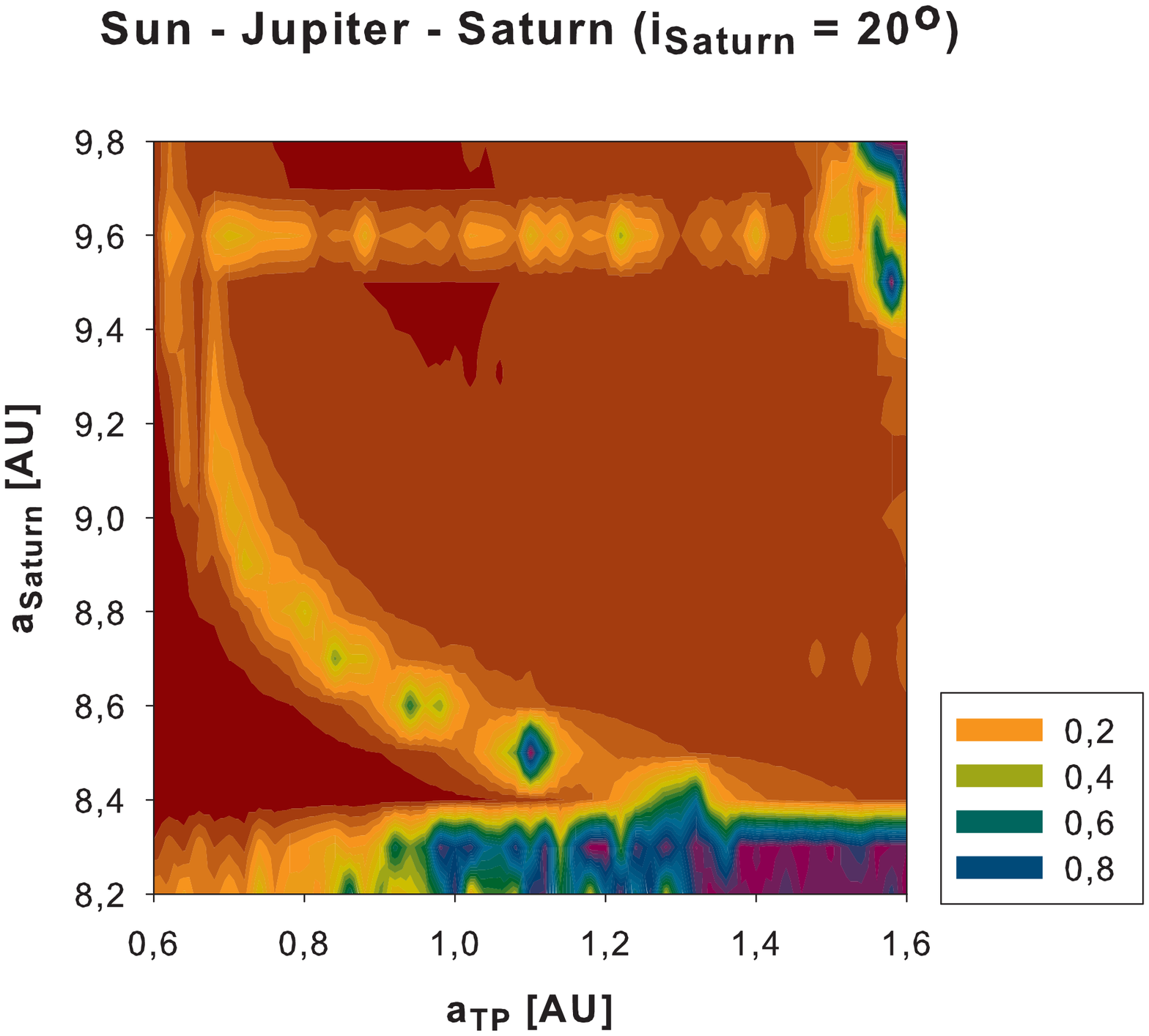}}
\centering{ 
\includegraphics[height=5cm]{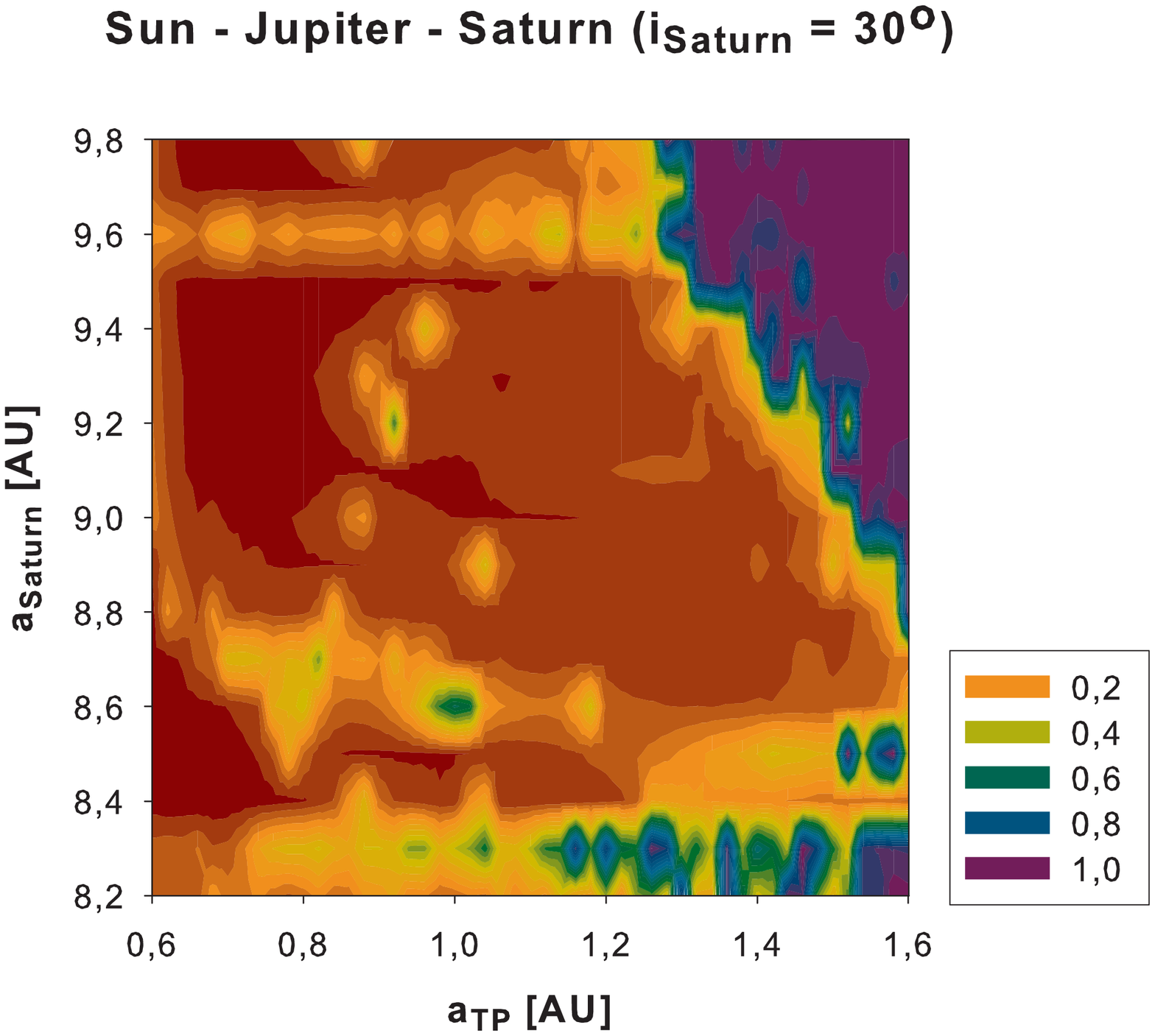}   
\includegraphics[height=5cm]{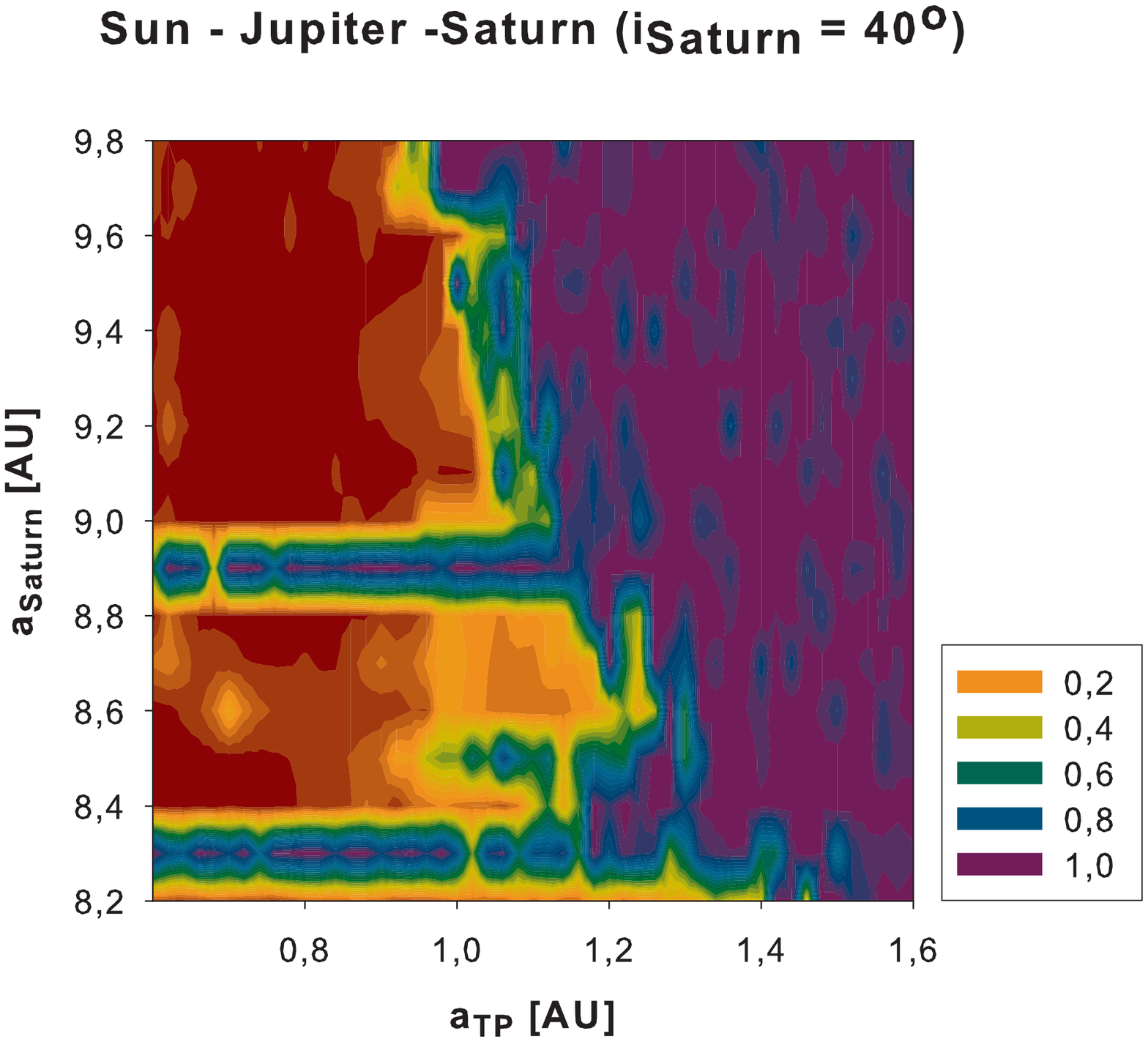} 
\includegraphics[height=5cm]{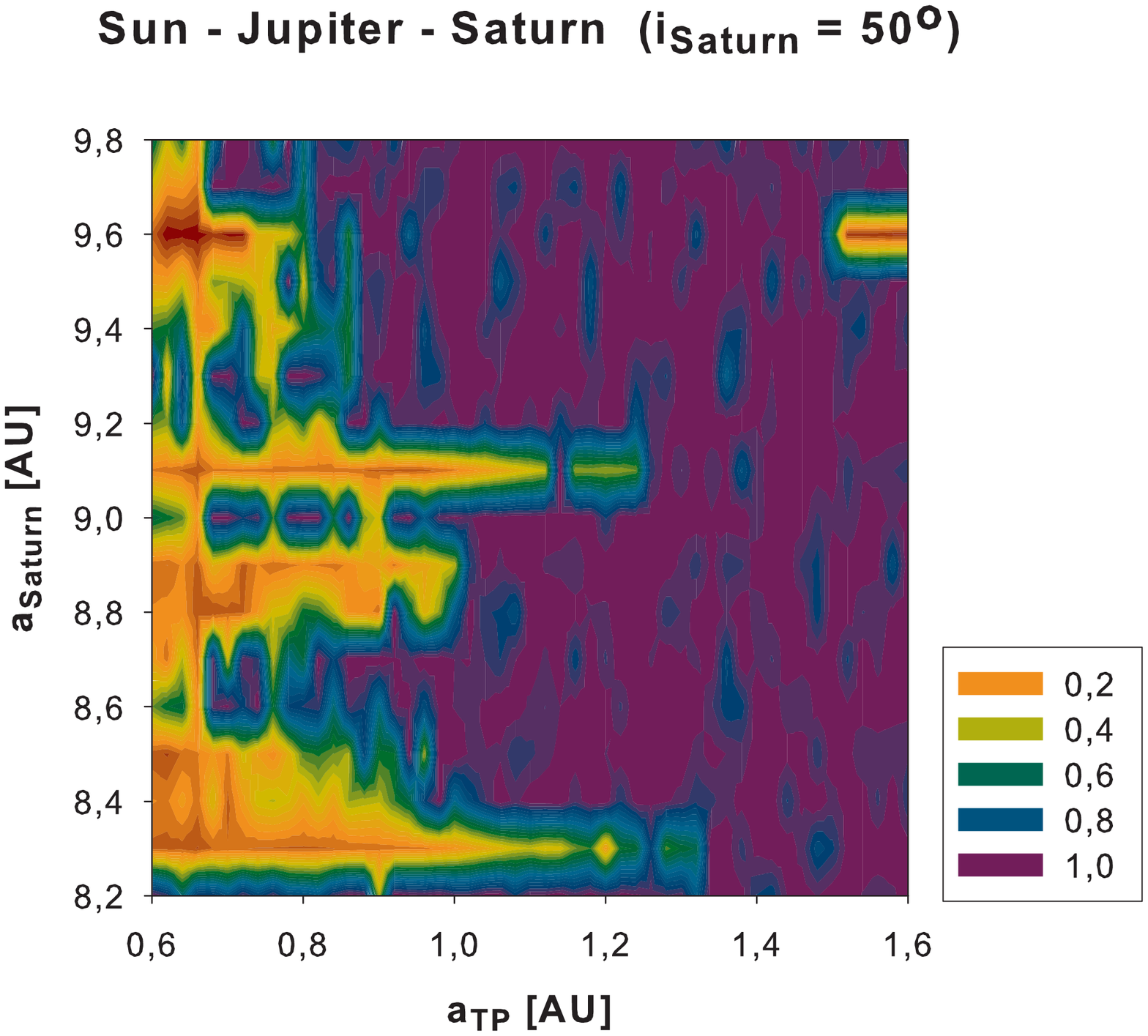}}
    
\caption[]{Max-e of test-planets in the HZ (x-axes) for
  different positions of Saturn (y-axes). Each map shows the maximum
  eccentricity for a certain inclination of Saturn: from the  planar
  case (top left panel) to $50^o$ (bottom right panel). Different colors
  denote different maximum eccentricities: from nearly circular (red areas) to
  unstable (purple).}
\label{epl-f10}
\end{figure}

\begin{figure}
\centering{
\includegraphics[height=8cm,angle=270]{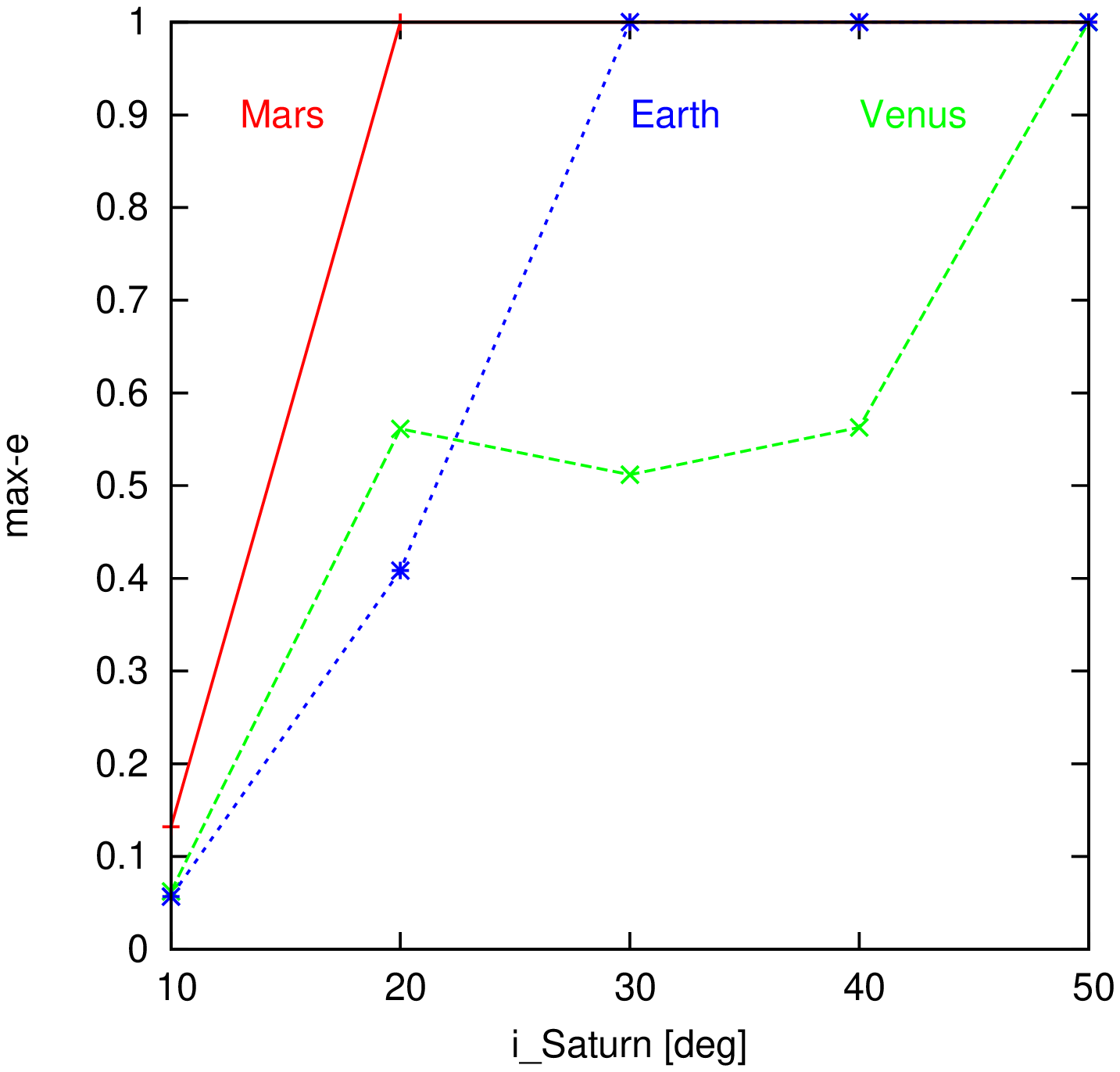}
}
\caption[]{Maximum eccentricities of the orbits of Venus, Earth and Mars in
  the different Jupiter-Saturn configurations where Saturn's orbits is
  inclined  between $10^o$ and $50^o$.}
\label{epl-f11}
\end{figure}

\section{Stability of Venus, Earth and Mars in inclined Jupiter-Saturn systems}

The max-e maps of Figs.~\ref{epl-f10}(a-f) allow also to determine the
stability of terrestrial planets Venus (at 0.72 au), Earth (at 1 au) and
Mars (at 1.52 au) when Saturn is at its actual position. The panels of
various inclinations of Saturn's orbit show that
first perturbations appear for Mars when $i_{\rm Saturn}=20^\circ$. 
Mars’ orbit would then be in yellow-green area
of Fig.~\ref{epl-f10}c (left panel in the middle) which corresponds to
max-e values between 0.2 and 0.3. 
In that case, Mars’ aphelion distance would be between 1.82 au and 1.98 au
which is in the unstable area. For Saturn’s inclinations $\geq 30^\circ$, 
the semi-major axis of Mars  is in the unstable area (purple region) which 
would lead immediately to an escape of this planet from the system. However, 
before Mars escapes, it will perturb the orbits of
Earth and Venus and as a result, the dynamical behavior of all three
planets change. 

Numerical computations of the actual orbits of Venus, Earth and Mars
in the various inclined  
Jupiter-Saturn configurations using the  parameters of the Solar system for
the planets Venus through Saturn confirmed this. A summary of
these computations is shown in Fig.~\ref{epl-f11}, where the maximum
eccentricities of Venus, Earth and Mars are plotted for different inclinations
of Saturn (from $10^o$ to $50^\circ$). The result shows that only for an
inclination of $10^\circ$  of Saturn's orbit the eccentricities of all three
planets (Venus-Mars) remained small.
For higher inclinations, one or several planets escaped from the system
and the remaining ones have high eccentricities. In the case of $i_{\rm Saturn}=50^\circ$, 
all three terrestrial planets escaped from the system.

\section{Conclusion}

In this paper, we studied the dynamical behavior of test-planets
moving in the water-based HZ of Sun-Jupiter-Saturn like systems. The HZ
was defined to be the area between 0.95 au and 1.37 au and the
test-planets  were assumed to be terrestrial planets having  similar
conditions as Earth.

In the first part of this paper, we analyzed a particular
configuration for which we knew from a previous study (PL08a) that a
test-planet at 1 au would show strong variations in its eccentricity. This
is the case when Jupiter orbits the Sun at 5.2 au and Saturn's semi-major axis
is changed to 8.7 au instead of 9.53 au.
This Jupiter-Saturn configuration perturbs the HZ and leads to variations
in eccentricity between 0 and nearly 0.7 for a test-planet at 1 au due to a secular frequency 
associated with the precession of Jupiter's perihelion. 
In the case of an eccentric orbit, the planet might leave
the HZ periodically which also depends on the size of the HZ.
If the HZ is defined as the area between 0.97 au and 1.37 au (according to
the works by  Kasting et al., 1993 and Kopparpou et al., 2013a \& b) the
test-planet will exit the HZ at peri-astron for eccentricities $> 0.03$
and at apo-astron when $e>0.4$. For a high-eccentric motion with
$e=0.7$, only 23\% of the planet’s orbit will be inside the HZ.  
If we consider a larger HZ, i.e.\ between 0.95 au  and 1.7 au (according to
the studies by Leconte et al. 2013 and Forget \& Pierrehumbert 1997)
then 70\% of the highly eccentric orbit would be inside the HZ. 

In our study, we also checked the influence of the ice planets(Uranus and 
Neptune) and of the neighboring planets (Venus and Mars) where we showed that the latter ones
are more important as they can decrease the eccentricity at 
1 au, while Uranus and Neptune have no significant influence. 

The second part of this paper analyzed the role of the mutual inclination 
of the two giant planets. We showed that moderate relative
inclinations (up to $30^\circ$) would decrease the secular perturbation in the HZ
and consequently the maximum eccentricities. In contrast, areas
affected by MMRs of Jupiter and Saturn are more perturbed in the inclined
systems. For higher inclinations (up to $40^\circ$) only the motion in the inner part of
the HZ is stable while in the outer part it is completely chaotic. 
An increase of Saturn's inclination enlarges this chaotic area. 

Finally, we studied the actual orbits of Venus, Earth and Mars 
in the various inclined Jupiter-Saturn configurations using the Solar
system  parameters (except for Saturn's inclination). 
These computations showed that for the architecture of the inner Solar
system to remain stable, the inclination of Saturn's orbit can increase only up to 
about $10^\circ$. For inclinations $\geq 20^\circ$,  we observed
escapes of one or several of the three terrestrial planets and high
eccentricity motion for the remaining planets.
An inclination of Saturn of $50^\circ$ for Saturn’s orbit removed all terrestrial planets
from the system.

This study shows that the planets of the Solar
system have to move nearly in the same plane to ensure low-eccentricity orbits
for the terrestrial planets which is probably necessary for the habitability
of Earth.

{\bf Acknowledgments}
The author would like to thank the Austrian Science Fonds FWF for financial support
of this work (projects P22603-N16 and S11608-N16).

\noindent 
{\bf References}
\bigskip \\
Bazso, A., Dvorak, R., Pilat-Lohinger, E., Eybl, V., Lhotka, Ch.: 2010, {\it
  CeMDA}, {\bf 107}, 57
\medskip \\
Chambers, J.E.: 1999, {\it MNRAS}, {\bf 304}, 793
\medskip \\
Eggl, S., Pilat-Lohinger, E., Georgakarakos, N., Gyergyovits, M., Funk, B. :
2012, {\it ApJ}, {\bf 752}, 74
\medskip \\
Eggl, S., Pilat-Lohinger, E., Funk, B., Georgakarakos, N., Haghighipour, N.:
2013, {\it MNRAS}, {\bf 428}, 3104
\medskip \\
Forget, F., Pierrehumbert, R.T.: 1997, {\it Science}, {\bf 278}, 1273
\medskip \\
Froeschl\'e, C.: 1984, {\it CMDA}, {\bf 34} 
\medskip \\
Froeschl\'e, C., Lega, E., Gonczi, R.: 1997, {\it CMDA}, {\bf 67}, 41
\medskip\\
Gehman, C.S., Adams, F.C., Laughlin, G.: 1996, {\it PASP}, {\bf 108}, 1018
\medskip \\
Jones, B.W., Sleep, P.N.  : 2002,  {\it A\&A}, {\bf 393}, 1015
\medskip \\
Jones, B.W., Underwood, D.R., Sleep, P.N.:2005, {\it ApJ}, {\bf 622}, 1091
\medskip \\
Kasting, J.F., Whitmire, D.P., Reynolds, R.T.: 1993, {\it Icarus},
{\bf 101}, 108. 
\medskip\\
Kopparapu, R.K., Ramirez, R., Kasting, J.F., Eymet, V., Robinson, T.D,
Mahadevan, S., Terrien, R.C., Domagal-Goldman, S., Meadows, V., Deshpande R.:
2013a, {\it ApJ}, {\bf 765}, 131
\medskip \\
Kopparapu, R.K., Ramirez, R., Kasting, J.F., Eymet, V., Robinson, T.D,
Mahadevan, S., Terrien, R.C., Domagal-Goldman, S., Meadows, V., Deshpande R.:
2013b, {\it ApJ}, {\bf 770}, 82
\medskip \\
Laskar, J. 1990, {\it Icarus}, {\bf 88}, 266
\medskip \\
Leconte, J., Forget, F., Charnay B., Wordsworth, R., Pottier, A.: 2013, {\it
  Nature}, {\bf 504}, 268
\medskip\\
Levison, H.F., Morbidelli, A., Tsiganis, K., Nesvorny, D., Gomes, R.: 2011,
{\it AJ}, {\bf 142}, 152
\medskip\\
Menou, K., Tabachnik, S.: 2003, {\it ApJ}, {\bf 583}, 473 
\medskip \\
Mischna, M.A., Kasting, J.F., Pavlov, A., Freedman, R.: 2000, {\it Icarus}, {\bf
  145}, 546 
\medskip \\
Morbidelli, A., Tsiganis, K., Crida, A., Levison, H.F., Gomes, R.: 2007, {\it
  AJ}, {\bf 134}, 1790
\medskip \\
Murray, C.D., Dermott, S.F. 1999, 
{\it Solar System Dynamics}, Cambridge University Press 
\medskip \\
Pilat-Lohinger, E., S\"uli, \'A., Robutel, P., Freistetter, F.: 2008a,
{\it ApJ}, {\bf 681}, 1639
\medskip\\
Pilat-Lohinger, E., Robutel, P., S\"uli, \'A., Freistetter, F.: 2008b,
{\it CeMDA}, {\bf 102}, 83
\medskip \\
Robutel, P., Gabern, F.: 2006, {\it MNRAS}, {\bf 372}, 1463 
\medskip  \\
S\'andor, Zs., S\"uli, A., \'Erdi, B., Pilat-Lohinger, E., Dvorak, R.: 2007, 
{\it MNRAS}, {\bf 375}, 11495  
\medskip \\ 
Tsiganis, K., Gomes, R., Morbidelli, A., Levison H.F.: 2005, {\it Nature},
{\bf 435}, 459
\medskip \\ 
Williams, D.M., Pollard, D: 2002, {\it IJAsB}, {\bf 1}, 61

\end{document}